\documentstyle[aps,epsfig,float,twocolumn,prl]{revtex}
\newcommand{\be}{\begin{equation}}
\newcommand{\ee}{\end{equation}}

\begin{document}
\twocolumn[\hsize\textwidth\columnwidth\hsize\csname @twocolumnfalse\endcsname
\draft
\title{Fractal to Nonfractal Phase Transition in the Dielectric Breakdown Model}
\author{M. B. Hastings}
\address{
CNLS, MS B258, Los Alamos National
Laboratory, Los Alamos, NM 87545, hastings@cnls.lanl.gov 
}
\date{8 March 2001}
\maketitle
\begin{abstract}
A fast method is presented for simulating the dielectric-breakdown model using
iterated conformal mappings.  Numerical results for the dimension
and for corrections to scaling
are in good agreement with the recent RG prediction of 
an upper critical $\eta_c=4$, at which a transition 
occurs between branching fractal clusters and one-dimensional nonfractal 
clusters.
\vskip2mm
\end{abstract}
]

Diffusion-limited aggregation\cite{dla}(DLA) is one of the most interesting
and difficult fractal growth models, despite its simple rules.  It models
a wide range of processes in nature that give rise to branching structures, 
including viscous fingering\cite{vf}, electrodeposition\cite{dep}, and 
dendritic growth\cite{dend}.  Yet, despite the importance of the model and
various theoretical approaches\cite{thy}, a full, controlled
understanding is lacking.

An extension of DLA is the dielectric-breakdown model\cite{dbm}(DBM), also
known as the $\eta$ model.  This model provides a continuously
varying fractal dimension $d$, ranging from 2 ($\eta=0$, the Eden 
model\cite{eden,eta0}), to roughly 1.7 ($\eta=1$, DLA), to 1.  
Recently\cite{rg}, it has been argued that $d=1$ for $\eta\geq \eta_c=4$, 
leading
to a controlled renormalization group expansion for the DBM.  
Further, while continuum DLA has certain integrability properties, most other
theoretical approaches do not rely on these and should apply to
the DBM as well.  Thus, numerical studies of the DBM provide a strong check
on {\it any} theoretical approach to fractal growth.  Given the intense recent
activity in the field, stimulated by the conformal mapping 
approach\cite{conf1,conf2,conf3,conf4}, such a check is crucial.

Several numerical studies of the DBM have been 
performed\cite{sim1,sim2,sim3}, and in one such study\cite{sim3} an
$\eta_c$ equal to 4 was suggested.  However, 
the difficulty of numerically solving Laplace's equation has restricted
the cluster sizes to roughly 5000 particles for $\eta=2$ and to
roughly 1000 particles for larger $\eta$.  We show below that the
numerically measured dimension of clusters with $\eta\geq 4$ 
decreases with increasing cluster size, and that 1000 particles is not
sufficient to extract correct information.

In this work, we present a fast algorithm for the DBM using
iterated conformal mappings, leading to an order of magnitude increase
in cluster size.  Unlike numerical solution of a discrete Laplace equation,
this technique is free of grid effects and boundary effects.
The numerical results are in good agreement
with upper critical $\eta$ found in the RG\cite{rg},
and for $\eta\geq \eta_c$, we find that the finite
size corrections to the numerically measured dimension are in good agreement
with the corrections predicted by the RG, and extrapolate well to $D=1$.

{\it Conformal Mapping Method ---} 
The conformal mapping technique for simulating DLA proceeds as follows:
define a function $F^{(n)}(z)$ that maps the unit circle in the complex plane
onto the boundary of the growing cluster after $n$ growth steps.  All
singularities of $F$ must lie within the unit circle, which implies that
$F^{(n)}(z)$ has a Laurent expansion
\be
F^{(n)}(z)=F_1^{(n)} z + F_0^{(n)} + F_{-1}^{(n)} z^{-1} + ...
\ee
Take $F^{(0)}(z)=z$.  Then, to obtain $F^{(n+1)}$ from $F^{(n)}$, pick a random
angle $\theta_{n+1}$, and construct an elementary mapping $f^{(n+1)}$ which 
produces a bump of linear size $\sqrt{\lambda_{n+1}}$ at angle $\theta_{n+1}$.
Then, compose mappings to define $F^{(n+1)}(z)=F^{(n)}(f^{(n+1)}(z))$, which
adds a bump to the cluster.  To obtain bumps of fixed size in the
physical plane, we chose
\be
\lambda_{n+1}=\frac{\lambda_0}{|F^{(n)'}(e^{i\theta_{n+1}})|^2},
\ee
where $\lambda_0$ is some constant.
Various elementary functions are possible; we use the function $f$
proposed in Ref.~\cite{conf1}, which includes a parameter $a$
describing the bump shape.

For the dielectric breakdown model with given
$\eta$, we need to chose the angle $\theta_{n+1}$ with probability 
proportional to $|F^{(n)'}(e^{i\theta_{n+1}})|^{1-\eta}$.  To do this, the 
program uses a Monte Carlo technique.  It starts with an angle
$\theta_{\rm old}$ which is initially taken equal to
$\theta_{n}$.  Then, it picks a random ``trial angle" $\theta$.
Define 
\be
P=\Bigl(\frac{|F^{(n)'}(e^{i\theta})|}{|F^{(n)'}(e^{i\theta_{\rm old}})|}
\Bigr)
^{1-\eta}.
\ee
If $P>1$, it replaces $\theta_{\rm old}$ with $\theta$, ``accepting"
angle $\theta$.  If $P<1$, it
replaces $\theta_{\rm old}$ with $\theta$ with probability $P$.
The program proceeds through a certain number $T$ of these trials, and
then takes $\theta_{n+1}=\theta_{\rm old}$ to add the $n+1$-st particle
to the cluster.  If $T$ is large enough, this yields the desired distribution.

The time for the algorithm to produce a cluster of size $N$ scales
as $N^2 T$.  We have used various means to determine the minimum value
of $T$ needed.  In Fig.~1, we show a plot of $F_1$ against number of
growth steps for single realizations with $\eta=4,N=10000$ and
$T=50,300,400,500$.  While the curve for $T=50$ is clearly different, the other
curves are all close, suggesting that $T=300$ suffices.  Determinations of the
dimension using $T=300$ are also indistinguishable from those using
$T=500$; however, to be careful, in all quantitative results given below,
we have chosen $T=500$, except for $\eta=2$, where $T=200$ was used.
On an 850-Mhz Pentium III, a 10000 particle cluster with $T=500$ can
be simulated in roughly 33 hours.

To provide an analytical
estimate of the minimum $T$ needed we consider the worst case, a one-dimensional
cluster, obtained by a suitably regularized version of the mapping
$F(z)=F_1 z+F_1/z$.  For $\eta>2$, the probability distribution
$|F^{(n)'}(e^{i\theta})|^{1-\eta}$ diverges near $\theta=0,\pi$.
The appropriate regularization of the cluster at a physical
length of order $\sqrt{\lambda}$ cuts off this divergence
at an angle that scales as $\sqrt{\sqrt{\lambda}/F_1}\propto\sqrt{1/n}$.  
In order to have sufficient trials that the algorithm is able to find
a point in this narrow region, one must have $T>\sqrt{n}$.  For clusters
which are not one-dimensional, so that $F'(z)$ is less singular, a
smaller value of $T$ may be used.
A final check on the value of $T$ needed is to look at the number
of times a trial angle is accepted, as shown in Table I.

In all simulations we used $\lambda_0=0.5$.  We chose $a=2/3$, 
suggested by Davidovitch et.~al.~\cite{conf2} as minimizing numerical error 
due to regions in the fjords of the cluster in which the derivative of $F(z)$ 
varies rapidly.  An improved technique for dealing with these regions involving
an acceptance window was suggested by Stepanov and Levitov\cite{conf3}.  
However, for $\eta>1$, we automatically avoid regions with large
$|F'(z)|$, so that we do not expect to have to worry about this 
problem; this justifies the very simple plotting technique employed to 
produce the images of the clusters below.  We chose to plot only the images of 
certain points on the unit circle chosen to lie near singularities of the 
mapping, each point corresponding to a single growth step.

We used a parallel computer to generate many realizations of
each cluster.  However, generation of a single large cluster is
a process that parallelizes very efficiently.
Each node picks random angles and
computes the Jacobian, returning the results to a central process, which
does the acceptance calculation, and returns the chosen angle to the other
nodes.  This will increase cluster size, but at a cost in statistics.

{\it Results ---}
In Fig.~2, we show a large cluster with $\eta=2,N=50000,T=30$ for 
purposes of illustration.  In Figs.~3,4 we show
clusters with $\eta=3.5,4.5$ and $N=10000,T=500$.  A difference in the
``fuzziness" of the two clusters is clear, with the cluster at $\eta=3.5$ having
many more small side branches.  The single side branch in Fig.~4 is not
the only possibility for this $\eta$.  Other realizations show either a cluster 
with no side branches, or with a single branching near a growth tip.

For quantitative results, we have generated 13-14 realizations of clusters 
for each $\eta=3,3.5,4,4.5,5$,
with $N=10000,T=500$, and 8 realizations of clusters for
$\eta=2$ and $N=15000,T=200$.
$F_1^{(n)}$ provides a measure of the linear size of the 
cluster; it may be shown that the radius of the cluster is at most
4 times $F_1$.  We calculated the dimension $D$ by averaging $F_1$ over
realizations of clusters and using linear regression on a log-log plot to fit
\be
F_1^{(n)} \propto n^{1/D}.
\ee
In Fig.~5 we plot $F_1$ and the rms fluctuations in $F_1$ 
for $\eta=3$ on a logarithmic scale; the rms fluctuations have a smaller 
slope, so the relative fluctuations in $F_1$
tend to zero, as found previously for $\eta=1$\cite{conf2}.  
The use of $F_1$ to determine dimension avoids problems with finite
size effects in box-counting that can lead to a spurious dimension less than 1,
as found for some $\eta$ in previous studies\cite{sim3}.

While for $\eta=3$, Fig.~5 is close to straight,
for $\eta=4$ there is a curvature, as shown in the solid line in
Fig.~6 (the dashed line is a fit discussed below).
To quantitatively measure this curvature, indicative of
corrections to scaling, 
we performed the linear regression 4 times for each $\eta$, first on
the full set, then on the last half, quarter, and eighth of the data set.
The results are shown in
Table II; we also show the rms fluctuations in $F_1$ after the
$10000$-th growth step.  
For $\eta<4$, the corrections to scaling are small, and the
dimension is relatively constant, sometimes increasing and sometimes decreasing
in the last part of the data set; these dimensions are close to
those found in previously\cite{sim1,sim2,sim3}.  

For $\eta\geq 4$, there is a clear trend for the dimension to
approach unity at longer times, with the trend most clear for $\eta>4$.  
We test the trend against the RG prediction for the corrections to 
scaling.  The leading irrelevant variable in the RG\cite{rg} is a tip splitting
rate $g$, which is irrelevant for $\eta>4$, with scaling dimension $(4-\eta)/2$.
At $\eta=4$ the tip-splitting is marginal, which may be shown to lead to
$F_1=c_1 n/(\log{(n/c_2)})^{c_3}$, where $c_3$ is a universal constant
determined by the $\beta$-functions and $c_2$ is nonuniversal.  
The dashed line in Fig.~6 shows the fit at
$\eta=4$ with this scaling form; the fit is indistinguishable from
the data over more than three decades, with only a slight difference
on the first 5 data points out of 10000.  (Within the RG, the
$\beta$-functions are determined numerically, so we 
keep $c_3$ as a fitting parameter; the fit agrees
with the numerical result from the RG).
For $\eta>4$ the leading correction to scaling is
$F_1=c_1 n (1+(n/c_2)^{(4-\eta)/2})$.  A fit with this form is shown in
Fig.~7 for $\eta=4.5$, 
with the fit distinguishable from the data over only roughly the
first 30 data points out of 10000.
(We checked the RG prediction for the scaling corrections by attempting
to fit the numerical data with other forms; the results are
much less accurate).

{\it Conclusion ---}
We have presented a fast algorithm for studying the dielectric breakdown
model.  Numerical results using this algorithm, using the
RG corrections to scaling, support an upper critical
$\eta_c=4$ for which the DBM clusters become one-dimensional.
The existence of finite size corrections for $\eta\geq 4$ makes it
desirable to simulate even larger clusters, but, while we have greatly
improved the cluster size over previous work, the results in this paper
are close to the limit of what can presently be 
attained using the conformal mapping techniques.

{\it Acknowledgements ---}
I would like to thank T. C. Halsey, L. Levitov, F. Levyraz, P. Pfeifer,
and I. Procaccia for useful discussions, as well as the Centro Internacional 
de Ciencas in Cuernavaca for a very interesting recent conference on
DLA and other nonlinear systems.  Numerical simulations were performed using the
Avalon computer at Los Alamos National Laboratory, as well as desktop machines.
This work was supported by DOE grant W-7405-ENG-36.

\begin{table}
\label{table1}
\caption{For various values of $\eta,N,T$, the total number of acceptances,
and number of growth steps on which at least one
acceptance occurred, for a typical run.
}
\begin{tabbing}
\hskip4mm \= $\eta$ \hskip 7mm \=  $N$  \hskip 7mm \= $T$
\hskip7mm \= Total \hskip12mm \= Growth Steps \\
\> \> \>            \>   Acceptances      \> with Acceptance \\
\> 2 \> 15000 \> 200 \> 1291738 \> 15000 \\
\> 2 \> 50000 \> 30 \> 580805 \> 49902 \\
\> 3 \> 10000 \> 500 \> 837291 \> 10000 \\
\> 3.5 \> 10000 \> 500 \> 442854 \> 10000 \\
\> 4 \> 10000 \> 500 \> 226170 \> 9990 \\
\> 4.5\> 10000 \> 500 \> 143031 \> 9950 \\
\> 5 \> 10000 \> 500 \> 99702 \> 9826 \\
\end{tabbing}
\end{table}
\begin{table}
\label{table2}
\caption{For various values of $\eta$, calculated dimension based on
different parts of data set; also, rms fluctuations in $F_1$ after last step,
normalized by $F_1$.
}
\begin{tabbing}
\hskip4mm \= $\eta$ \hskip 6mm \=  full set  \hskip 6mm \= last 1/2
\hskip6mm \= last 1/4 \hskip6mm \= last 1/8 \hskip 4mm \= rms \\
\> 2 \> 1.433 \> 1.4256 \> 1.435 \> 1.4522 \> .039 \\
\> 3 \> 1.263 \> 1.264 \> 1.262 \> 1.243 \> .056 \\
\> 3.5 \> 1.170 \> 1.143 \> 1.155 \> 1.162 \> .078 \\
\> 4 \> 1.128 \> 1.090 \> 1.078 \> 1.071 \>  .072\\
\> 4.5 \> 1.101 \> 1.090 \> 1.066 \> 1.035 \> .039 \\
\> 5 \> 1.068 \> 1.030 \> 1.025 \> 1.009 \> .046 \\    
\end{tabbing}
\end{table}
\begin{figure}[!t]
\begin{center}
\leavevmode
\epsfig{figure=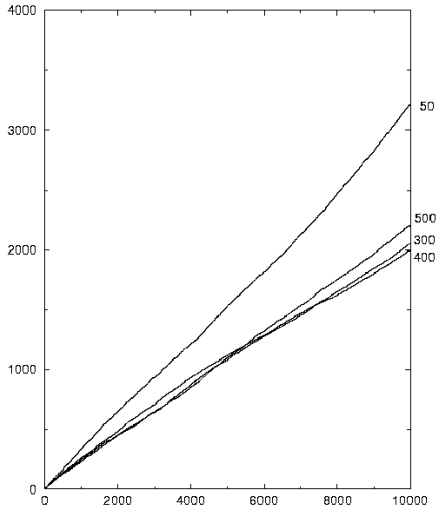,height=9cm,angle=0,scale=.75}
\end{center}
\caption{$F_1$ as a function of $N$ for $\eta=4$ and various values of $T$
(marked on the right-hand axis).}
\label{4rel}
\end{figure}
\begin{figure}[!t]
\begin{center}
\leavevmode
\epsfig{figure=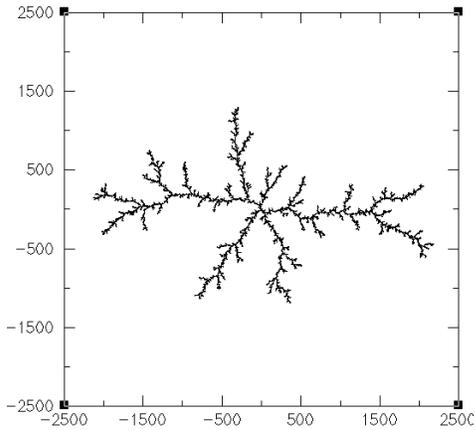,height=8cm,angle=0,scale=.75} 
\end{center}
\caption{Cluster with $\eta=2,N=50000,T=30$.}
\label{fig1}
\end{figure}
\begin{figure}[!t]
\begin{center}
\leavevmode
\epsfig{figure=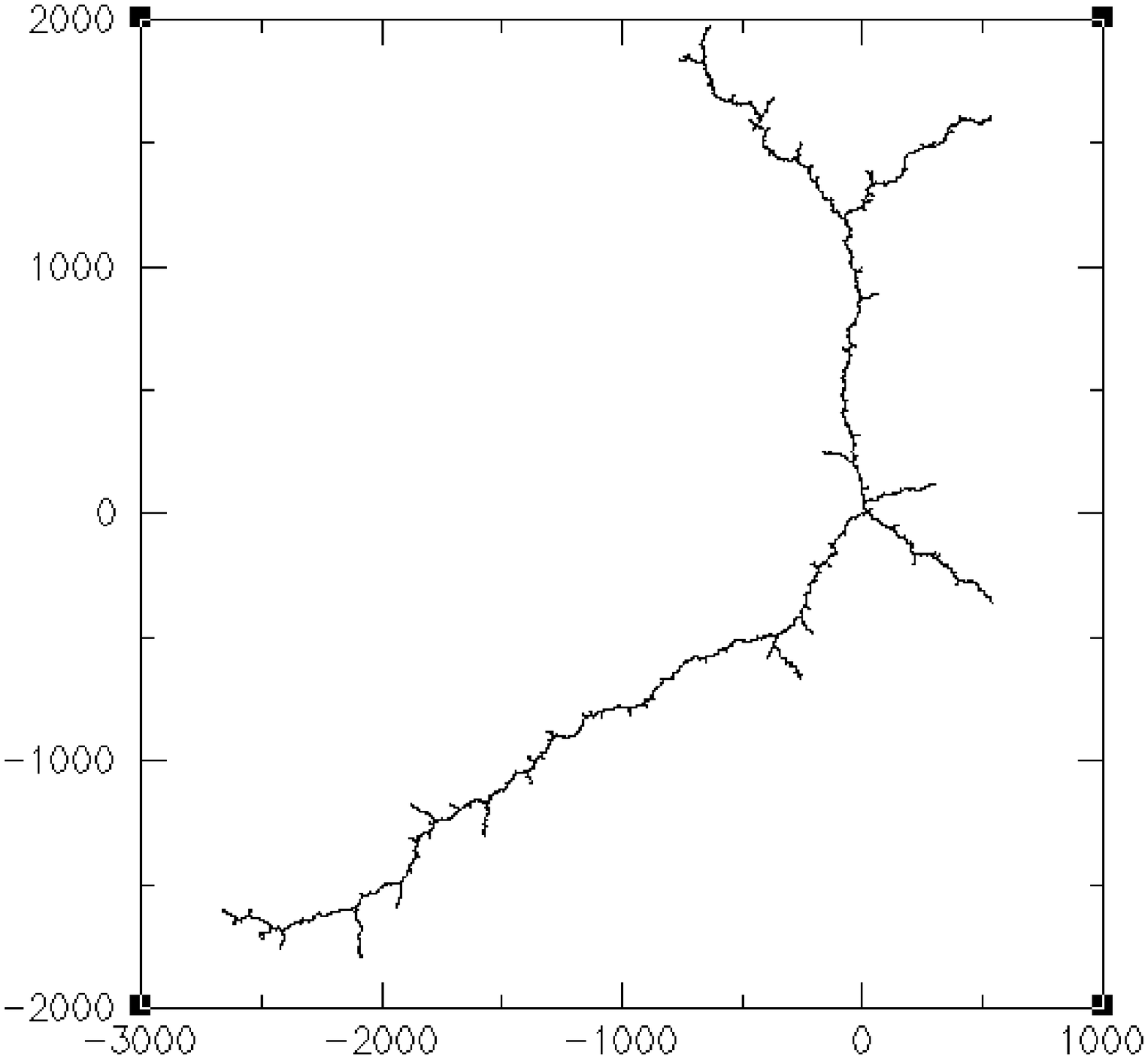,height=8cm,angle=0,scale=.75}
\end{center}
\caption{Cluster with $\eta=3.5,N=10000,T=500$.}
\label{fig3}
\end{figure}
\begin{figure}[!t]
\begin{center}
\leavevmode               
\epsfig{figure=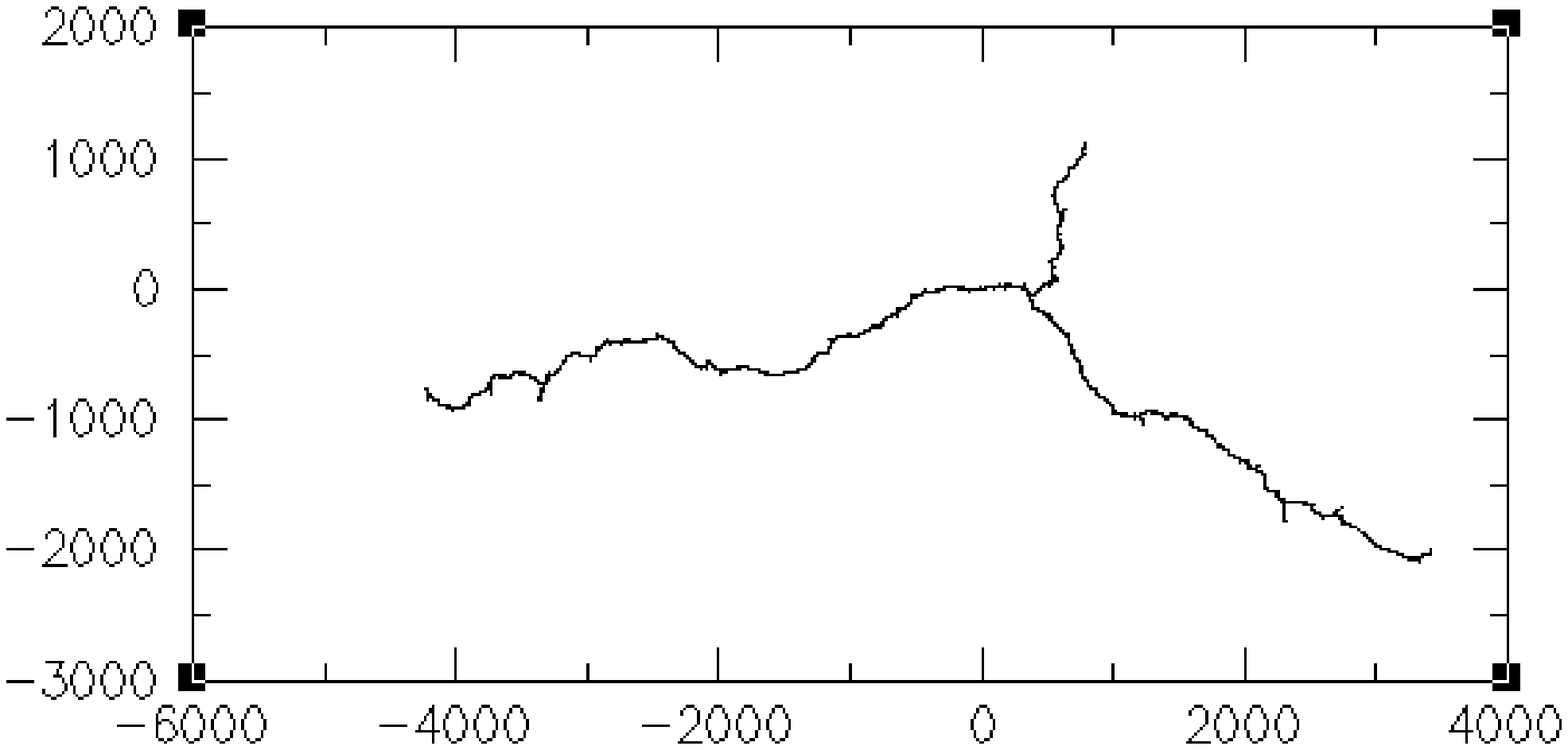,height=6cm,angle=0,scale=.75}
\end{center}
\caption{Cluster with $\eta=4.5,N=10000,T=500$.}
\label{fig4}
\end{figure}
\begin{figure}[!t]
\begin{center}
\leavevmode
\epsfig{figure=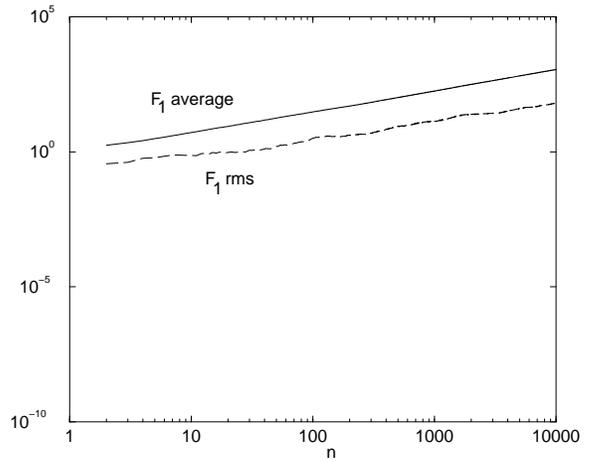,height=8cm,angle=0,scale=.75}
\end{center}
\caption{Average of $F_1$ (upper curve) and rms fluctuations in $F_1$ 
(lower curve) as a function of $n$ for $\eta=3$.}
\label{avd}
\end{figure}
\begin{figure}[!t]
\begin{center}
\leavevmode
\epsfig{figure=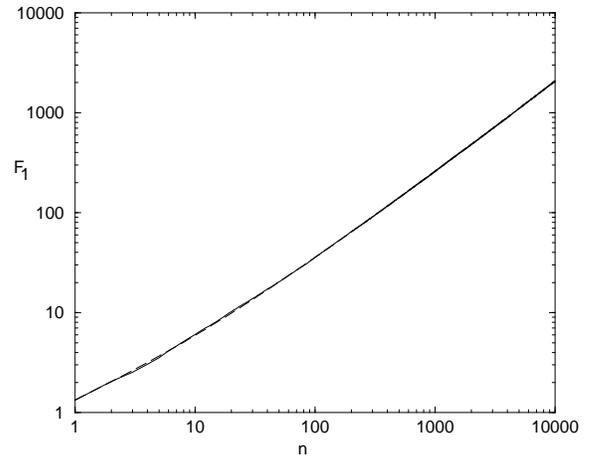,height=8cm,angle=0,scale=.75}
\end{center}
\caption{Plot of $F_1$ versus $n$ for $\eta=4$.  The barely
visible dashed line is a fit including corrections to scaling.}
\label{llc}
\end{figure}    
\begin{figure}[!t]
\begin{center}
\leavevmode
\epsfig{figure=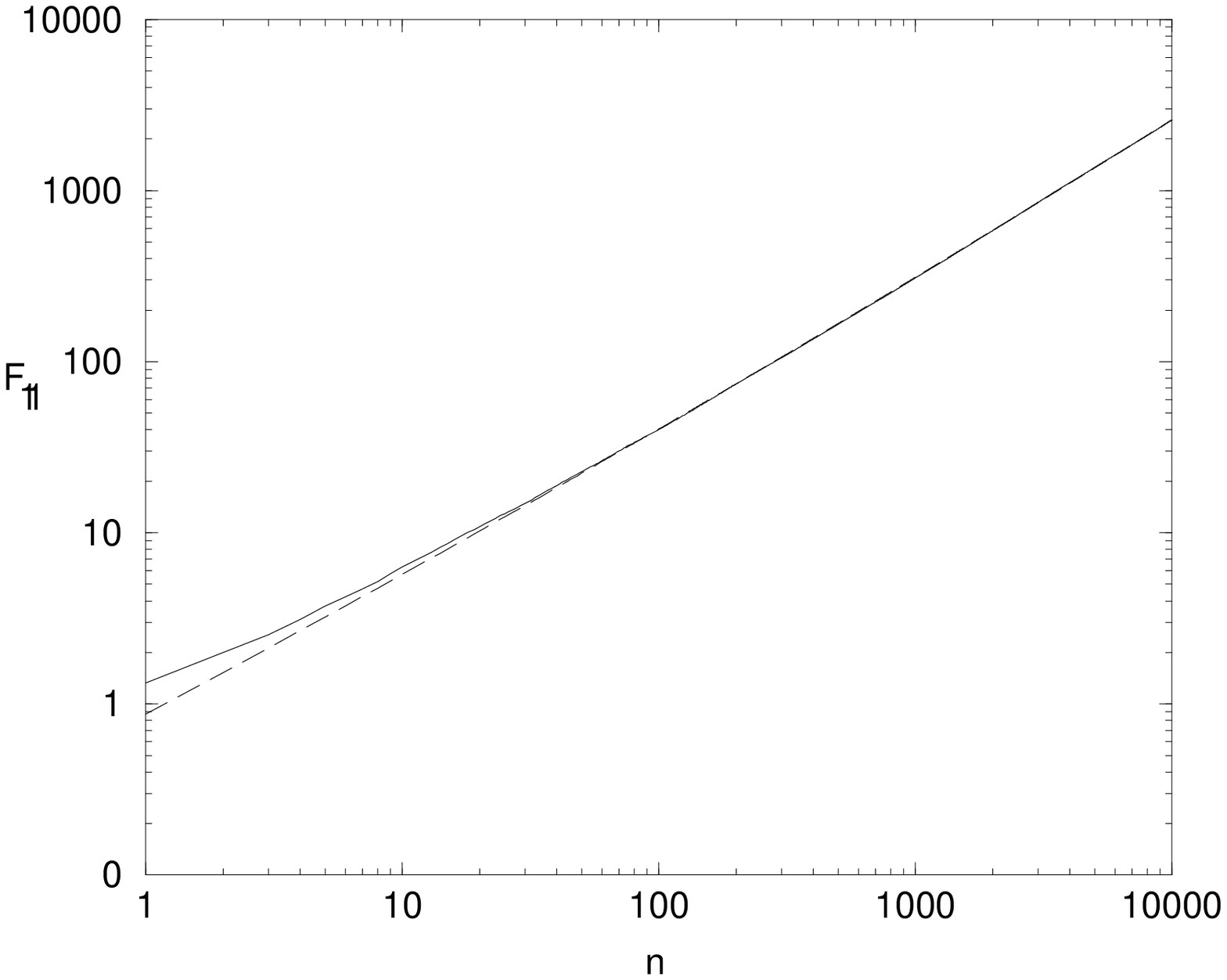,height=8cm,angle=0,scale=.75}
\end{center}
\caption{Plot of $F_1$ versus $n$ for $\eta=4.5$.  The dashed line is
a fit including corrections to scaling.}
\label{llc2}
\end{figure}    


\begin{thebibliography}{99}
\bibitem{dla}  T. A. Witten and L. M. Sander, Phys. Rev. Lett. {\bf 47}, 1400
(1981).

\bibitem{vf} J. Nittman, G. Daccord, and H. E. Stanley,
Nature {\bf 314}, 141 (1985).

\bibitem{dep} D. Grier, E. Ben-Jacob, R. Clarke, and L. M. Sander, 
Phys. Rev. Lett. {\bf 56}, 1264 (1986); R. M. Brady and R. C. Ball, Nature 
{\bf 309}, 225 (1984).

\bibitem{dend} J. Kert\'esz and T. Vicsek, J. Phys. A {\bf 19}, L257 (1986).
 
\bibitem{dbm}  L. Niemeyer, L. Pietronero, and H. J. Wiesmann, Phys. Rev. Lett.
{\bf 52}, 1033 (1984).

\bibitem{thy} B. Davidovitch, A. Levermann, and I. Procaccia, Phys. Rev. E 
{\bf 62}, R5919 (2000); T. C. Halsey, Phys. Rev. Lett {\bf72}, 1228 (1994);
A. Erzan, L. Pietronero, and A. Vespignani, Rev. Mod. Phys. {\bf 67}, 545 
(1995).

\bibitem{eden} M. Eden, in Proc. 4-th Berkeley Symposium of Mathematics, 
Statistics, and Probability, Ed. F. Neyman (University of California Press, 
Berkeley, 1961).
        
\bibitem{eta0} M. B. Hastings, preprint cond-mat/9910274.

\bibitem{rg} M. B. Hastings, preprint cond-mat/0104344.

\bibitem{conf1} M. B. Hastings and L. S. Levitov, Physica D {\bf 116}, 244
(1998).

\bibitem{conf2} B. Davidovitch et.~al., Phys. Rev. E {\bf 59}, 1368 (1999).

\bibitem{conf3} M. G. Stepanov and L. S. Levitov, preprint cond-mat/0005456.

\bibitem{conf4} E. Somfai, L. M. Sander, and R. C. Ball,
Phys. Rev. Lett. {\bf 83}, 5523 (1999).    

\bibitem{sim1}  C. Amitrano, Phys. Rev. A {\bf 39}, 6618 (1989).

\bibitem{sim2} C. Evertsz, Phys. Rev. A {\bf 41}, 1830 (1990).

\bibitem{sim3} A. Sanchez et.~al., Phys. Rev. E {\bf 48}, 1296 (1993).
\end{thebibliography}
\end{document}